\documentclass[ preprint, amssymb, amsmath, acs, jpc]{revtex4-1}

\usepackage{comment,bbold,dcolumn,csquotes,xspace,graphicx,amsmath,tikz,multirow,bbold}
\usepackage[nolist]{acronym}
\usepackage[normalem]{ulem}
\usepackage[utf8]{inputenc}
\usepackage{graphicx}
\usepackage{subcaption}
\usepackage{gensymb}
\usepackage{phoenician}
\usepackage{amsmath}

\usepackage[colorlinks=true,linkcolor=blue]{hyperref}%
\expandafter\ifx\csname package@font\endcsname\relax\else
 \expandafter\expandafter
 \expandafter\usepackage
 \expandafter\expandafter
 \expandafter{\csname package@font\endcsname}%
\fi
\usetikzlibrary{positioning}

\newcommand*{\citen}{}
\DeclareRobustCommand*{\citen}[1]{%
        \begingroup
        \romannumeral-`\x 
        \setcitestyle{super}%
        \cite{#1}%
        \endgroup
}

\newcommand{\gj}[1]{{\textit{} #1}}

\newcommand{\br}[0]{{\boldsymbol{r}}}
\newcommand{\bomega}[0]{{\boldsymbol{\Omega}}}
\newcommand{\kb}{\text{k}_\text{B}}
\newcommand{\DJS}{D_\text{JS}}

\newcommand\mydots{\hbox to 0.9em{.\hss.\hss.}}

\newcommand{\alf}{\text{\phncfamily{\ARaleph}}}

\begin{document}

\title{The Angular Localization Function (ALF): a practical tool to measure solvent angular order with Molecular Density Functional Theory.}

\author{Ma\"iwenn Souetre}
\affiliation{Sorbonne Université, CNRS, Physico-Chimie des Électrolytes et Nanosystèmes Interfaciaux, PHENIX, F-75005 Paris, France}

\author{Benjamin Rotenberg}
\email{benjamin.rotenberg@sorbonne-universite.fr}
\affiliation{Sorbonne Université, CNRS, Physico-Chimie des Électrolytes et Nanosystèmes Interfaciaux, PHENIX, F-75005 Paris, France}
\affiliation{Réseau sur le Stockage Électrochimique de l'Énergie (RS2E), FR CNRS 3459, 80039 Amiens Cedex, France}

\author{Guillaume Jeanmairet$^*$}%
\email{guillaume.jeanmairet@sorbonne-universite.fr}
\affiliation{Sorbonne Université, CNRS, Physico-Chimie des Électrolytes et Nanosystèmes Interfaciaux, PHENIX, F-75005 Paris, France}
\affiliation{Réseau sur le Stockage Électrochimique de l'Énergie (RS2E), FR CNRS 3459, 80039 Amiens Cedex, France}

\begin{abstract}
Molecular density functional theory is a powerful technique for efficiently computing the spatially and orientationally dependent equilibrium density of a molecular solvent around an arbitrary solute. This density encodes the detailed solvent structure, but contains so much information that it is difficult to interpret comprehensively. Although spatial dependence is frequently analyzed through orientationally integrated number density, angular information remains poorly exploited. The present work addresses this gap by introducing a function that provides a local measure of the angular order: the Angular Localization Function (ALF), derived from the ideal free energy functional, which quantifies the entropy. We discuss the connections between  ALF and well known statistical functions. We illustrate the utility of ALF by discussing the solvent structure for three systems immersed in water: water as a solute, an octanol molecule, and three clay minerals (talc, fluorotalc and pyrophyllite) with small differences in their structure leading to subtle effects on their interactions with water. ALF provides information complementary to quantities such as the average polarization or charge density to characterize the local orientational distribution of solvent molecules around solutes and next to surfaces. It also offers a convenient visualization tool akin to the Electronic Localization Function (ELF) used to analyze bonding in quantum chemistry. 
\end{abstract}

\maketitle

\section{Introduction}
In the absence of external perturbation, a fluid composed of rigid classical bodies, \textit{i.e} atoms or molecules whose interactions can be described by a classical force-field, is homogeneous and isotropic: On average, particles are uniformly distributed and oriented. The application of an external perturbation causes the fluid to become inhomogeneous. At the surface of minerals or metals, water forms molecular layers, with oscillating density and orientation profiles before reaching the bulk behavior (typically within approximately 1~nm). Similarly, solvent molecules are organized around solutes in solvation shells, with a degree of ordering decreasing with the distance from the solute.

Several theoretical and simulation approaches exist to compute the structural and thermodynamic properties of such perturbed fluids. Methods based on an explicit sampling of the configuration space, such as Molecular Dynamics and Monte Carlo are extremely popular. While these methods are computationally demanding, they yield in principle exact results, provided that the sampling is sufficient. However, the computational cost required to converge the estimation of local properties can quickly become prohibitive, even using enhanced sampling techniques or recently introduced estimators with a reduced variance~\citen{borgis_computation_2013, de_las_heras_better_2018, coles_computing_2019, rotenberg_use_2020, coles_reduced_2021, renner_reduced-variance_2023}. In fact, the variance of estimators based on histograms of the position (and orientation) diverges as the size of the bins vanishes, \textit{i.e.} as the resolution of the grid used to sample local properties increases. For instance, if one assumes that each bin must be visited at least 100 times to achieve an acceptable signal-to-noise ratio, computing a one-dimensional density distribution  over a 10~\AA~range with a 0.1~\AA~resolution requires generating approximately $\sim10^4$~independent configurations.  This requirement escalates dramatically to $10^8$ configurations for a three-dimensional density map resolved on a 10 \AA$^3$  cube with the same resolution. Computing the orientational distribution of fluid particles on such a grid becomes practically infeasible, even though its first moment, the local average polarization, can be sampled with the same limitations as the number density~\citen{coles_computing_2019}.

Classical density functional theory (cDFT)\citen{Mermin-PR-65,Evans-AP-79} is a powerful alternative to particle-based simulations for the evaluation of such local observables. It allows one to compute energetic and structural properties of the fluid by a variational principle (minimizing a free energy functional) rather than by a tedious sampling of the phase space. A particular strength of the theory is that it directly provides a detailed picture of the liquid structure. Building on the excellent quality of fundamental measure theory for the hard-sphere fluid\citen{rosenfeld_free-energy_1989}, accurate free energy functionals have been proposed to describe a wide range of fluids including polymers, colloids or electrolyte solutions\citen{lowen_density_2002, marechal_inhomogeneous_2011, oleksy_wetting_2010}. Molecular density functional theory (MDFT) is a flavour of cDFT that is designed to describe the solvation of a molecular solute into a molecular solvent, such as water or acetonitrile \citen{jeanmairet_molecular_2013_2, zhao_molecular_2011, ding_efficient_2017}. In MDFT, the solvent is described by its density field, which depends on both spatial and angular coordinates. 

While the equilibrium solvent density is easily computed within MDFT, it is challenging to comprehend because visualizing data defined on a 6-dimensional space is difficult. This issue is typically addressed by projecting the density onto lower-dimensional representations. For example, the radial distribution function between the center of mass of the solvent and a given site of the solute is a one dimensional quantity that can be computed at any distance $r$ by integrating the density within a thin spherical shell of radius $r$. Other common  representations include slices of density in a given plane, as well as isosurfaces (\textit{i.e.} hypersurfaces defined by a common value of the density).  Similarly, analysing  orientational information requires some projections. A popular one is the first angular moment, which corresponds to the  polarization field in the case of dipolar solvent molecules~\citen{levesque_solvation_2012, jeanmairet_hydration_2014}. Another example are the projections of the density onto  rotational invariants. Such projections are helpful to solve the molecular Ornstein-Zernike equation\citen{blum_invariant_1972,blum_invariant_1972-1,blum_invariant_1973}, or its MDFT formulation\citen{ding_efficient_2017,jeanmairet_molecular_2026}. They can also explain the long-range orientational order observed in scattering experiments\citen{belloni_screened_2018}.  \gj{One can also mention the local version of the nematic order parameter $S_2$ used in the context of liquid crystals phase transition\cite{sliwa_local_2017,ztakahashi_molecular_2023}. }

In the context of solvation, several strategies have been proposed to estimate local thermodynamic quantities, in particular the entropy. Most methods rely on a spatial discretization using a 3D grid, but  differ in their entropy estimator. In Grid Cell Theory~\citen{gerogiokas_prediction_2014} and 3D-2PT~\citen{lin_two-phase_2003,persson_signatures_2017}, entropy is estimated using analytical particle-based models, with values accumulated over time and averaged within each voxel. Grid inhomogeneous solvation theory (GIST)~\citen{lazaridis_inhomogeneous_1998, nguyen_grid_2012, nguyen_molecular_2019} is based on an exact entropy expansion in terms of solvent $n$-body correlation functions, that are evaluated on a 3D grid. In practice, the expansion is often limited to the one-particle term. A recent approach proposed by Heinz and Grubm\"uller~\citen{heinz_computing_2020, heinz_permut_2021} is based on the mutual information expansion (MIE), a concept from information theory. Entropies are computed directly by sampling the configuration space, taking advantage of permutation symmetry. Both the MIE and GIST expansion theoretically converge to the exact entropy, but  in practice, the expansion is truncated after the second order at best. Moreover the expansion terms  of MIE and GIST do not coincide beyond the first order. It is worth noting that these studies often focus on computing entropy maps around biological objects, where the fluid is inherently inhomogeneous, and the first-order term is the first non-trivial contribution. In homogeneous systems, the first non trivial term is the second order-term, which can be computed from the radial distribution function~\citen{dyre_perspective_2018,klumov_two-body_2020}.  This quantity has been empirically related to dynamical properties such as viscosity or heat conductivity through Rosenfeld's excess-entropy scaling\citen{rosenfeld_relation_1977}. Finally, as pointed out by Persson~\citen{persson_note_2022}, the decomposition into spatially resolved entropy is not uniquely defined. Therefore, one should remain cautious with the interpretation of such quantities. 

In the present work, we introduce a measure of  the local orientational order/disorder of the solvent around a solute or next to an interface, which we call the  Angular Localization Function (ALF), by analogy with the Electron Localization Function (ELF) used in electronic DFT \citen{becke_simple_1990}. While the ELF connects electron localization to the local kinetic-energy density, the ALF quantifies the local orientational entropy. As ELF, ALF can conveniently be used for visualization purposes. Mathematically, ALF is the Kullback-Leibler (KL) divergence~\citen{martin_jensenshannon_2015} of the isotropic orientation distribution with respect to the actual orientation distribution. The KL divergence is a well-known measure of the similarity between distributions, and comparing with the isotropic distribution quantifies how the actual angular distribution differs from the isotropic one observed in the bulk reference fluid. \citen{martin_jensenshannon_2015}. While its physical interpretation is less straightforward, it possesses valuable properties, such  as being bounded and returning distinct values for vacuum and homogeneous regions. \gj{The measures introduced here are illustrated using MDFT, but they can also be applied to other classical DFT methods\citen{marechal_inhomogeneous_2011,simon_orientational_2025}  or simulation techniques that allow for the computation of a spatially and orientationally dependent density. The approach is also straightforwardly generalizable to mixtures of anisotropic particles, where ALF can be computed for each constituent species.}
Section~\ref{sec:theo} shortly reviews MDFT and presents the concepts introduced in the present work to analyze the local angular distribution, in particular the angular localization function. Section~\ref{sec:results} then illustrates the utility of ALF on several systems.

\section{Methods \label{sec:theo}}

\subsection{Molecular Density Functional Theory \label{subsec:mdft}}

We begin with a brief overview of Molecular Density Functional Theory (MDFT). In MDFT, the liquid is described by its density, $\rho(\br,\bomega)$, which depends on spatial and orientational coordinates $\br$ and $\bomega$, respectively. The classical Density Functional Theory (cDFT) framework guarantees the existence of a variational principle for the grand potential.\citen{Mermin-PR-65,Evans-AP-79} Specifically, there exists a unique intrinsic Helmholtz free energy functional  of the fluid density, which, combined with an external potential, yields the grand potential functional. The equilibrium solvent density is the density that minimizes this grand potential functional, which is then equal to the grand potential of the system. While the theory guarantees the existence of an exact Helmholtz free energy functional~\citen{DwaSch-PRE-11,jeanmairet_variational_2026}, its explicit form remains unknown. Consequently, practical calculations require the use of approximate functionals.


In MDFT, the working functional is defined as the difference between the grand potential functional and the grand potential of the homogeneous, unperturbed, fluid. It is usually decomposed as follows
\begin{equation}
\label{eq:Fsplit}
    F[\rho(\br,\bomega)]=F_\text{id}[\rho(\br,\bomega)]+F_\text{exc}[\rho(\br,\bomega)]+\iint \rho(\br,\bomega) V_\text{ext}(\br,\bomega){\rm d}\br {\rm d}\bomega
\end{equation}
The ideal term, $F_\text{id}$, corresponds to the free energy of the ideal gas, \textit{i.e.} without interactions between fluid particles, and will be discussed in more detail in Section~\ref{subsec:alf}. The excess term $F_\text{exc}$ account for these interactions, while the last integral arises from the interactions between the fluid particles and the external field, $V_\text{ext}$, describing the presence of the solute. The most challenging term is the excess functional, for which no practical exact expression exists. Thus, approximations are required and many have been proposed to describe hard-bodies\citen{rosenfeld_free-energy_1989, kierlik_density-functional_1991, roth_fundamental_2002, hansen-goos_density_2006}, charged systems\citen{roth_shells_2016, cats_capacitance_2022} or molecular liquids\citen{jeanmairet_molecular_2013_2, levesque_scalar_2012, zhao_molecular_2011, zhao_site-site_2013, borgis_simple_2020, borgis_accurate_2021}. The present work does not focus on evaluating the accuracy of MDFT predictions. Instead, its primary objective is to introduce the angular localization function (ALF) as a tool to analyze the orientational order of the solvent. The excess term is  approximated here by the HNC functional, which has been previously described\citen{ding_efficient_2017}. But the ideas introduced in the following can be applied with any free energy functional of the solvent position and orientation. \gj{Since using a different approximation for the excess functional will lead to different predictions of the equilibrium density, this will naturally affect ALF in turn. In fact, ALF predictions from reference simulations could be used, alongside other quantities such as radial distribution functions or solvation free energies, to guide the construction of the excess functional.}

\subsection{Angular Localization Function \label{subsec:alf}}

We now introduce a measure of the local angular ordering encoded in the distribution $\rho(\br,\bomega)$. Our starting point is the ideal functional,
\begin{equation}
\label{eq:Fid}
\beta F_\text{id}[\rho(\br,\bomega)]=\iint f_\text{id}(\rho(\br,\bomega)){\rm d}\br {\rm d}\bomega
\end{equation}
with
\begin{equation}
\label{eq:fid}
f_\text{id}(\rho(\br,\bomega))=\rho(\br,\bomega)\ln\frac{\rho(\br,\bomega)}{\rho_{0}}-(\rho(\br,\bomega)-\rho_{0}).
\end{equation}
In Eq~\ref{eq:Fid}-\ref{eq:fid} {$\beta=(\kb T)^{-1}$}, where $\kb$ is the Boltzmann constant, $T$ is the temperature and {$\rho_0=n_0/8 \pi^2$} where $n_0$ is the homogeneous number density of the fluid and the $1/8 \pi^2$ factor accounts for the isotropic angular distribution. This ideal functional measures the entropic cost, for a non interacting fluid, to depart from the homogeneous and isotropic density $\rho_{0}$ to reach the distribution $\rho(\br,\bomega)$. It is worth emphasizing that this term does not account for the total entropy, since the excess term $F_\text{exc}$ includes entropic contributions arising from multi-body correlations.

Since the entropy is a measure of (dis)order in a thermodynamic system, we will start from $F_\text{id}$ to build a quantity that measures the local angular order. To that end, we first split the  density $\rho$ into a spatial part and an angular part, 
\begin{equation}
\label{eq:alphadef}
\rho(\br,\bomega)=n(\br)\alpha(\br,\bomega)
\end{equation}
with the number density
\begin{equation}
\label{eq:ndef}
n(\br) \equiv \int\rho(\br,\bomega){\rm d}\bomega \; ,
\end{equation}
so that by construction the angular part $\alpha(\br,\bomega)\equiv \rho(\br,\bomega)/n(\br)$ is normalized over orientations for each $\br$ :
\begin{equation}
\label{eq:alphanorm}
\int\alpha(\br,\bomega){\rm d}\bomega=1 \; .
\end{equation}
For the isotropic distribution, $\alpha(\br,\bomega)=\alpha_0=1/8\pi^2$.

We now introduce a position-dependent dimensionless entropy function by integrating the corresponding free energy density Eq.~\ref{eq:fid} over orientations at fixed position $\br$: 
\begin{equation}
\label{eq:S(r)}
S(\br) \equiv \int  f_\text{id}(\rho(\br,\bomega)){\rm d}\bomega \, .
\end{equation}
Using the definition Eq.~\ref{eq:alphadef}, this leads to:
\begin{align}
S(\br) &= \int \left[ n(\br)\alpha(\br,\bomega)\left(\ln\frac{n(\br)}{n_{0}}+
\ln\frac{\alpha(\br,\bomega)}{\alpha_{0}}\right)-n(\br)\alpha(\br,\bomega)+\rho_{0}\right] {\rm d}\bomega 
\nonumber \\
&= n(\br)\ln\frac{n(\br)}{n_{0}}-n(\br)+n_{0} + n(\br)\int\alpha(\br,\bomega)\ln\frac{\alpha(\br,\bomega)}{\alpha_{0}}{\rm d}\bomega
\end{align}
where we used the normalization Eq.~\ref{eq:alphanorm} and the fact that $\rho_0=n_0\alpha_0$.
This can be rewritten as 
\begin{equation}
\label{eq:defSnSomega}
S(\br)=S_{n}(\br)+S_{\Omega}(\br)
\end{equation}
where
\begin{equation}
\label{eq:Sn}
S_{n}(\br)\equiv n(\br)\ln\frac{n(\br)}{n_{0}}-n(\br)+n_{0}
\end{equation}
depends only on the number density and
\begin{equation}
\label{eq:Somega}
S_{\Omega}(\br) \equiv n(\br)\int\alpha(\br,\bomega)\ln\frac{\alpha(\br,\bomega)}{\alpha_{0}}{\rm d}\bomega
\end{equation}
accounts for the local orientational disorder.
While $S_{n}(\br)$ quantifies the entropic cost for the number density to deviate from the average density $n_0$, $S_{\Omega}(\br)$ is a measure of the angular anisotropy. However, $S_{\Omega}(\br)$ is proportional to the local number density. Consequently, regions with strong preferential orientation but low number density can have a value of $S_{\Omega}$ comparable to those of poorly oriented region with high number density. In order to discriminate between such situations to focus on the angular distribution, we define the angular localization function (ALF), \gj{denoted with the eponymous Phoenician letter, $\alf$, as}
\begin{equation}
\label{eq:ALF}
\alf(\br) \equiv \frac{S_{\Omega}(\br)}{n(\br)} 
= \int \alpha(\br,\bomega) \ln\frac{\alpha(\br,\bomega)}{\alpha_0}{\rm d}\bomega \; .
\end{equation}
This definition corresponds to the Kullback–Leibler (KL) divergence $D_\text{KL}(\alpha_0||\alpha)(\br)$ of the isotropic distribution $\alpha_0=1/8\pi^2$ with respect to the actual density $\alpha(\br,\bomega)$ \citen{vedral_role_2002}. Thus, $\alf$ is necessarily positive and  vanishes for the isotropic distribution ($\alpha=\alpha_0$). In the continuous case, it is not bounded above and can diverge, for instance, for a Dirac distribution. In the discrete case, it is maximized by the perfectly oriented distribution where all but one orientation have a zero angular density. Alternatively, it is also possible to define the KL divergence of $\alpha(\br,\bomega)$ with respect to $\alpha_0$:
\begin{equation}
\label{eq:KL}
D_\text{KL}(\alpha||\alpha_0)(\br)=\int\alpha_0\ln\frac{\alpha_0}{\alpha(\br,\bomega)}{\rm d}\bomega \;.
\end{equation}
While this expression might seem a mathematically simpler choice to quantify the deviation from the isotropic distribution, it diverges when $\alpha(\br,\bomega)\to0$, which can occur for some $\bomega$ in highly oriented regions, even in the discrete case. Since ALF does not suffer from this problem and follows more naturally from physical quantities, as described above, it will be preferred to the KL divergence of Eq.~\ref{eq:KL}. Finally, we introduce a third quantity, the Jensen-Shannon divergence\citen{martin_jensenshannon_2015}: 
\begin{equation}
\label{eq:JS}
D_\text{JS}(\br)=\frac{1}{2}\int \left[\alpha_0\ln\left(\frac{2\alpha_0}{\alpha(\br,\bomega)+\alpha_0}\right)+\alpha(\br,\bomega)\ln\left(\frac{2\alpha(\br,\bomega)}{\alpha(\br,\bomega)+\alpha_0}\right) \right]{\rm d}\bomega.
\end{equation}
With respect to the KL divergences of Eqs.~\ref{eq:ALF}-\ref{eq:KL}, the Jensen-Shannon divergence offers the advantage of being symmetric in $\alpha(\br,\bomega)$ and $\alpha_0$. It vanishes when $\alpha=\alpha_0$ and is bounded (by $\ln(2))$.

As mentioned in the introduction, several approaches have been developed to compute local entropies using molecular simulations. The local entropies $S_{n}$ and $S_{\Omega}$ directly corresponds to the first-order terms  of GIST\citen{nguyen_grid_2012}, which also identifies with the first-order terms of the mutual information expansion, used for example in the Per$|$Mut method~\citen{heinz_computing_2020, heinz_permut_2021}. While these methods were originally developed for MD simulations, they have also been applied with liquid state theories. For instance, Nguyen \textit{et. al.}  used  3D-RISM  to compute  entropy maps around  a protein, specifically the coagulation Factor Xa\citen{nguyen_molecular_2019} based on GIST\citen{lazaridis_inhomogeneous_1998}. However, since RISM adopts an atomic site rather than a molecular description in the Ornstein-Zernike equation, the local entropies $S_{n}$ and $S_{\Omega}$ cannot be computed directly. Instead, molecular distributions needs to be approximately reconstructed from site distribution, introducing errors in the computed density.


\section{Results and Discussion \label{sec:results}}

We now illustrate the capabilities of the introduced quantities, ALF and $D_\text{JS}$, to characterize the orientational order of the solvent around a solute. Specifically, we consider
three systems immersed in water: water as a solute (Section~\ref{subsec:water}), an octanol molecule  (Section~\ref{subsec:octanol}), and three clay minerals (talc, fluorotalc and pyrophyllite) with small differences in their structure leading to subtle effects on their interactions with water  (Section~\ref{subsec:clays}).

\subsection{Water as a solute \label{subsec:water}}

\begin{figure}[ht!]
    \begin{center}
            \includegraphics[width=\columnwidth]{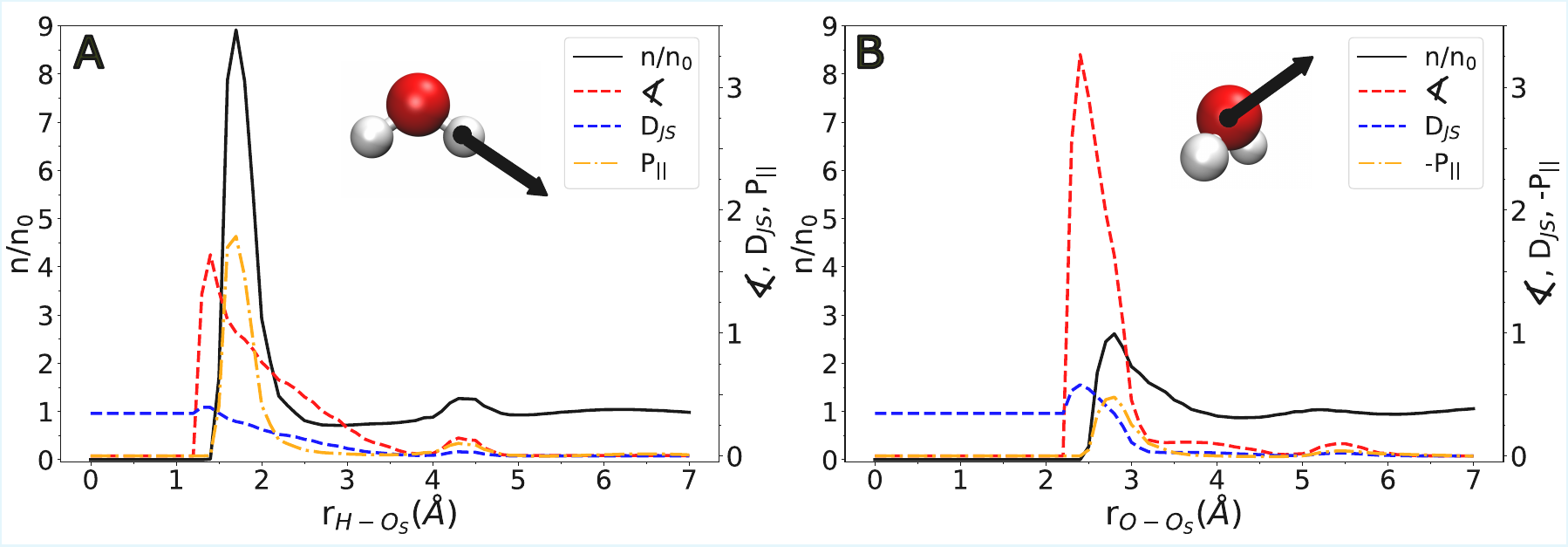}
    \end{center}
    \caption{
    Oxygen density with respect to the homogeneous density $n(\br)/n_0$ (solid black), angular localization function $\alf(\br)$ defined by Eq.~\ref{eq:ALF} (dashed red), Jensen-Shannon divergence $\DJS$ defined by Eq.~\ref{eq:JS} (dashed blue) and parallel polarization (dotted orange) computed according to Eq.~\ref{eq:P_parallel} around a water molecule. Results are computed from the equilibrium density $\rho(\br,\bomega)$ obtained by MDFT and plotted along two specific axes (indicated by black arrows) around a water molecule: (A) along an O-H bond and (B) along an oxygen lone-pair direction (see text). The distance is \gj{measured between the oxygen atom of the solvent and} the H or O atom \gj{of the solute} in the former and latter case, respectively. \gj{Note that these two distances are shifted by 1~\AA~which is the length of an OH bond in the SPC/E model}. The number density $n/n_0$ is a dimensionless quantity whose values can be read on the left axis.  $\alf$ and $\DJS$ are dimensionless quantities whose values can be read on the right axis. The polarization density is in arbitrary units to fit on the graph and since the sign of the projection $P_\parallel$ in the direction considered in panel (B) is negative, we report $-P_\parallel$ in that case.
    }
    \label{fig:water_along_OH_lonepairs}
\end{figure}

We start by considering the solvent structure around a water molecule in liquid water described by MDFT. The solute water is described by the SPC/E forcefield and the solvent functional also employs the direct correlation functions\citen{belloni_exact_2017} and forcefield parameters of SPC/E.
The density was computed within a $25\times25\times25$~\AA$^3$ box discretized on 250$^3$ grid points, with 196 discrete orientations used to describe the rotation space spanned by the three Euler angles~\citen{ding_efficient_2017}. Fig.~\ref{fig:water_along_OH_lonepairs} reports several quantities computed from the equilibrium density $\rho(\br,\bomega)$, along 2 particular axes: through an O-H bond (panel A), and along an oxygen to lone-pair direction panel (B). More precisely, the second direction is obtained from the former via $S_4$ improper rotation about the $C_2$ axis of the molecule. Results are reported as a function of the distance \gj{between the oxygen atom of the solvent and} the H or O atom \gj{of the solute} in panels~\ref{fig:water_along_OH_lonepairs}a and~\ref{fig:water_along_OH_lonepairs}b, respectively. Specifically, we consider the oxygen density normalized by its bulk value,  $n/n_0$ (black {solid} line), $\alf$ (red {dashed} line) and $\DJS$ (blue {dashed} line), as well as the projection on the considered axis of the average local dipolar polarization density ({orange dotted} line), 
\begin{equation}
\label{eq:P_parallel}
    P_\parallel(r)={\bf P}(\br)\cdot{\bf u}
\end{equation}
 with ${\bf u}$ a unit vector in the relevant direction.
 The dimensionless dipolar polarization is defined as
 \begin{equation}
     {\bf P}(\br)=\int \rho(\br,\bomega) \bomega {\rm d}\bomega.
\end{equation}
Since the sign of the projection $P_\parallel$ in the direction considered in Fig.~\ref{fig:water_along_OH_lonepairs}b is negative, we report $-P_\parallel$ in that case. The polarisation is in arbitrary units to be displayed on the same graph. 

The number density and parallel component of the polarization are consistent with the expected tetrahedral structure around the water molecule. Along the O-H bond (Fig.~\ref{fig:water_along_OH_lonepairs}a), the density peak at $\approx2$~\AA~from the solute's hydrogen and the positive $P_\parallel$ with a peak at the same position indicate the presence of a solvent molecule receiving a H-bond from the solute. In contrast, in the direction of the lone pair (Fig.~\ref{fig:water_along_OH_lonepairs}b), the number density and negative parallel polarization are consistent with a molecule donating an H-bond, with the oxygen of the solvent molecule at $\approx3$~\AA~from the solute's oxygen. The lower value of the maximum of $|P_\parallel|$ in the latter case is partly due to the fact that the dipole of the solvent molecule is not aligned with the O-H bond donating the H-bond to the solute, whereas in the former case it is aligned with the O-H bond of the solute donating the H-bond to the solvent.


Since $\alf$ and $\DJS$ are not usual quantities, we begin with general considerations. By convention, we set $\alpha(\br,\bomega)=0$ when $n(\br)=0$, causing $\alf$ to vanish both in regions with zero density (very close to the solute) and in regions where all orientations are equivalent such as in the homogeneous and isotropic region far from the solute (see Fig.~\ref{fig:water_along_OH_lonepairs}). For its part, the Jensen-Shannon divergence, distinguishes between these 2 regions: it vanishes in the bulk solvent but reaches a plateau in the solvent-free region. In the intermediate region, where the solvent is inhomogeneous and anisotropic, the two functions behave very similarly. This suggests that both quantities can be used interchangeably to quantify the orientational order. For the remainder of this discussion, we will focus on ALF.

Along the O-H bond (Fig.~\ref{fig:water_along_OH_lonepairs}a), \gj{ALF} reaches a maximum closer to the solute's hydrogen than the density peak, its value then decreases gradually toward zero. Several conclusions on the local solvent order can be drawn from this profile. The peak in ALF corresponds to regions with strong preferential orientation but very low number density. Indeed, as the solvent approaches the solute, both the steric repulsion (captured by the Lennard-Jones site on the oxygen in the SPC/E model) and the electrostatic interactions are stronger.  Consequently, while molecules are excluded by the steric repulsion (small $n$), those that are present are strongly oriented by the large electric field. ALF quantifies the narrowing of the angular distribution at short distance from the solute's OH. At longer distance, ALF decays much more slowly with $r$ than $n$ or $P_\parallel$, indicating the gradual broadening of the angular distribution (due to the decay of the electric field) of molecules present in this region -- a feature less visible on the polarization, which also depends on the amount of molecules present. A signature for the second solvation shell also appears around 4~\AA~in the density, parallel polarization and \gj{ALF}.
 
In the direction of the oxygen lone pair (Fig.~\ref{fig:water_along_OH_lonepairs}b), the peak of ALF is also closer to the solute than the peaks of the number and polarization densities. However the value of the maximum is significantly larger than along the OH bond of the solute, reflecting a larger deviation from the isotropic distribution, \textit{i.e.} a narrower distribution of orientations. Solvent molecules donating a H-bond can mainly rotate around the corresponding O-H bond (the small oscillation of the bond around its preferred orientation, called libration, corresponds to a limited range of orientations), whereas the ones receiving a H-bond from the solute can rotate around their bisector (the direction of the dipole) aligned with the donating O-H bond, but also other axes (leading to a misalignment of the dipole with the O-H direction). Here again, we note that ALF provides insights different from the polarization density, which depends on both the amount of molecules present and their configuration with respect to the considered axis. Interestingly, ALF decays much faster in this direction than along the O-H bond of the solute, with a distribution of orientations close to isotropic already near the farther end of the first solvation shell (which is also broader than in the other direction).

\subsection{Octanol in water \label{subsec:octanol}}

\begin{figure}[ht!]
    \centering
    \includegraphics[width=0.85\textwidth]{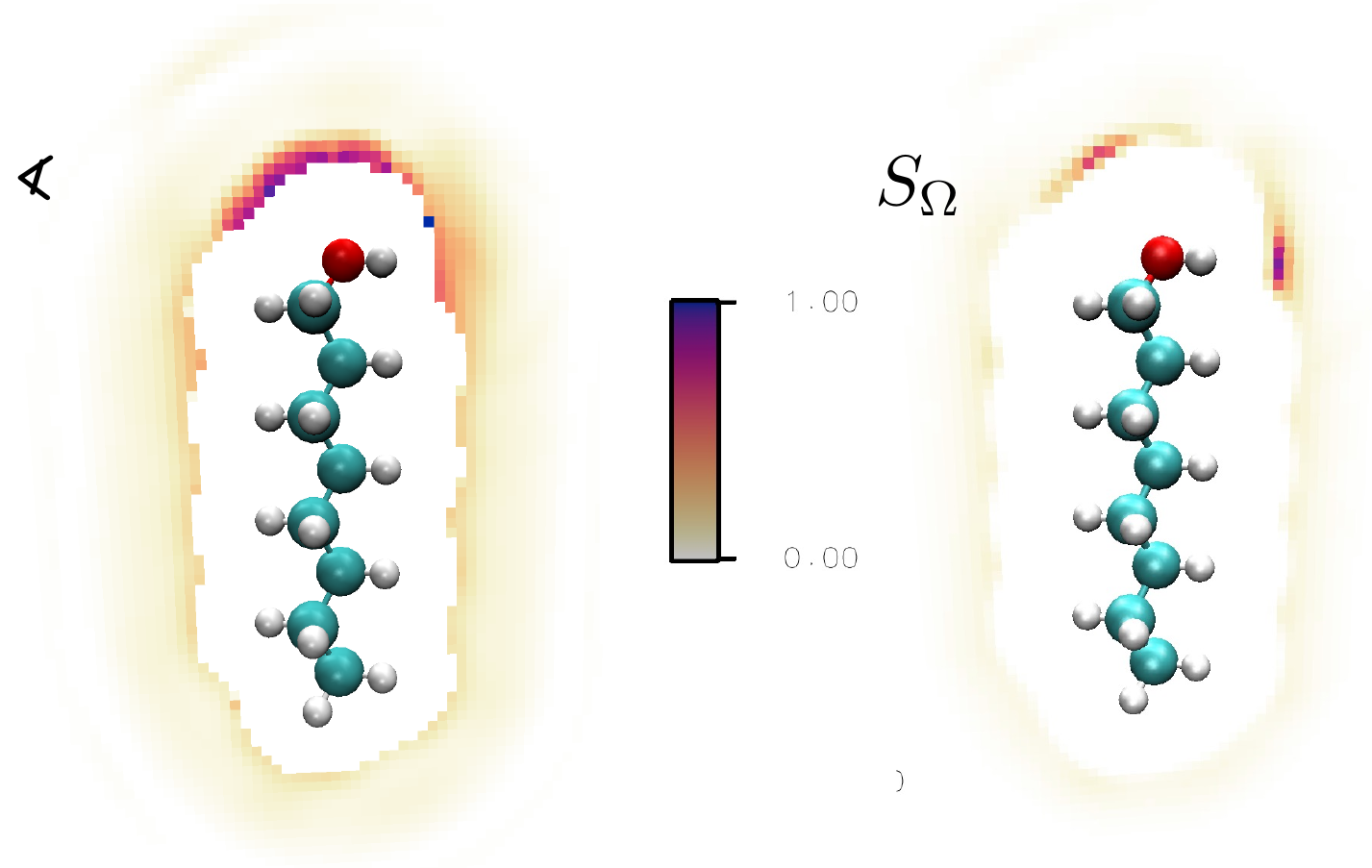}
    \caption{Slices of $\alf$ (left) and of $S_{\Omega}$ (right) obtained for SPC/E water solvent in the plane containing the C (cyan), O (red) and H (white) atoms of an octanol molecule using MDFT. Both quantities have been normalized with respect to the maximum value to ease comparison. High value regions are in magenta while zero regions are in white.}
    \label{fig:octanol_maps}
\end{figure}

We now turn to the properties of water around an octanol molecule, which contains different chemical moieties. The geometry and forcefield parameters are given in SI. We use the  same box size, grid resolution and number of angles as for the water case (see Section~\ref{subsec:water}). Fig.~\ref{fig:octanol_maps} first reports slices of $\alf$ and of the orientational entropy $S_{\Omega}$, defined in Eq.~\ref{eq:Somega}, in the plane containing the C, O and H atoms of octanol.  The overall picture  of the orientational entropy is consistent with the first order entropy computed using Per\textbar Mut reported by Heinz \textit{et al.} (see Fig.~4 of Ref.~\citenum{heinz_permut_2021}). However, the region of high entropy predicted with simulations appears more diffuse than the one predicted with MDFT. This quantitative difference can have several origins, such as insufficient sampling in the simulations, differences in the force fields used or limitations of free energy functional used in MDFT functional. 

The largest values of $\alf$ are observed close to the octanol O atom, while that of $S_{\Omega}$ are found next to the H atom. In addition, the basins of large $\alf$ values are wider than the corresponding ones of $S_{\Omega}$. These differences can be rationalized by noticing that, unlike $S_{\Omega}$, $\alf$ is not weighted by the density number $n$ (see Eq.~\ref{eq:ALF}). As a result, $\alf$ is a better measure of the orientational disorder of water molecules actually present at a given position around octanol. The water molecules have marked preferential orientation in the vicinity of the oxygen atom, even in regions where few water molecules are present. In contrast, near the OH bonds, even if the molecules are globally oriented, their angular distribution is more isotropic.  It is also worth noting that the density prefactor in  $S_{\Omega}$ affects the shape of the non-zero region surrounding the solute.  In  $S_{\Omega}$,  the extent of this region is limited compared to $\alf$, reflecting the fact that regions with the most inhomogeneous orientational distributions tend to have very low density. In the vicinity of the alkyl chain, the orientation distribution is more isotropic, as indicated by the low values of ALF and $S_{\Omega}$. \gj{Regarding the isolated bright pixel near the O-H bond, a detailed analysis of the contribution to ALF reveals that this point corresponds to a very low density, $n \approx 10^{-5}n_0$,  with $n_0$ the homogeneous bulk density. Its angular distribution vanishes everywhere except for one angle, which explains the very high value of ALF. Reducing the number of discrete angles used for the angular discretization yields an ALF slice almost indistinguishable from Fig. 2 but without this bright pixel. This  illustrates that in these low density regions, $\rho(\bm{r},\bm{\Omega})$ is higly sensitive to the discretization parameters.}

To conclude the discussion about the density maps, it is worth recalling that MDFT does not suffer from sampling issues inherent to MD simulations\citen{hsu_electron_2022}. Thus, a map of ALF as the one depicted in Fig~\ref{fig:octanol_maps}, can be obtained with MDFT at a computational cost reduced by several orders of magnitude with respect to a similar map of first order entropy computed with  MD.

\begin{figure}[t]
    \centering
    \includegraphics[width=\columnwidth]{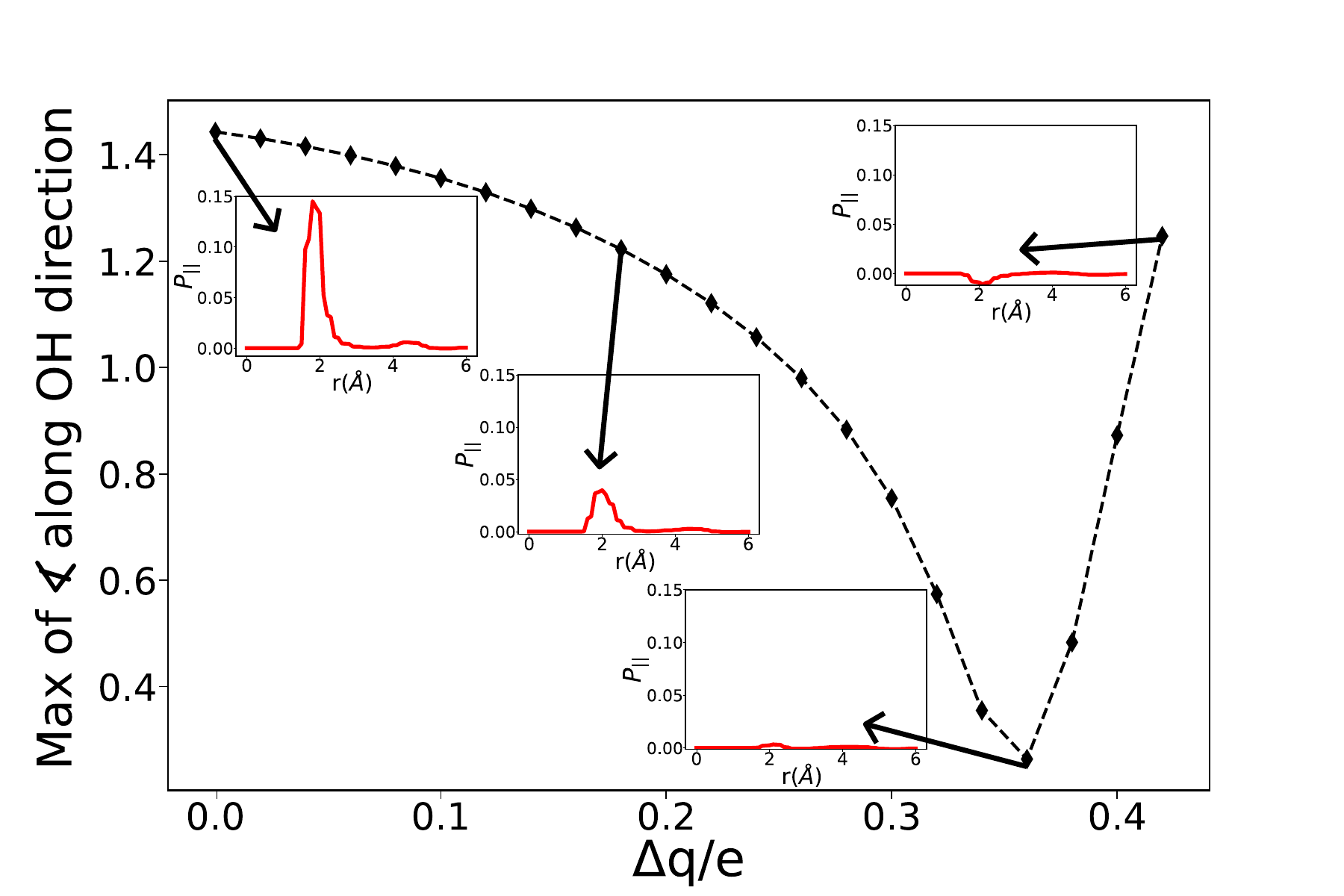}
    \caption{
    Maximum value of ALF along the O-H direction, for octanol molecules with a modified charge distribution within the alcohol group ($q_{\rm O}=q_{\rm O}^{ref}+\Delta q$ and $q_{\rm H}=q_{\rm H}^{ref}-\Delta q$). $\Delta q/e=0$ corresponds to the reference case discussed in Fig.~\ref{fig:octanol_maps}.     The parallel polarization profiles in the O-H direction, $P_\parallel(r)$, with $r$ the distance from the H atom are also shown as insets for some representative values of $\Delta q$. }
    \label{fig:max_ALF_as_of_deltaq}
\end{figure}

To investigate the impact of the polarization of the O–H bonds on the orientation of surrounding water molecules, we systematically reduced the O–H dipole moment by adding a  charge $\Delta q$ on the oxygen atom while withdrawing it from the hydrogen atom. Fig.~\ref{fig:max_ALF_as_of_deltaq} reports the maximum value of ALF along the O-H bond direction, for several values of $\Delta q$. This maximum decreases monotonically to reach a minimum at $\Delta q^*=0.36e$ (with $e$ the elementary charge) before increasing. In other words, the angular distribution becomes more isotropic as $\Delta q$ increases until $\Delta q^*$, while angular ordering is progressively restored beyond this value.
This can be rationalized by examining the parallel polarisation profiles in the O-H direction, $P_\parallel(r)$, with $r$ the distance from the H atom, which are displayed as insets Fig.~\ref{fig:max_ALF_as_of_deltaq} for selected values of $\Delta q$. 

For low values of $\Delta q$, the parallel polarization is positive, indicating an orientation of the corresponding water molecules consistent with an H-bond donated by the octanol. As $\Delta q$ increases, the polarity of the O-H bond and the corresponding electric field are reduced, hence the orientational disorder increases and ALF is reduced. This trend persists until the local polarization vanishes, not because there are no molecules at that position but because their average orientation vanishes, which is also reflected in the minimum of ALF. Beyond that point, the orientation of the local electric field is reversed and the polarization changes sign. As the magnitude of the field increases by further increasing $\Delta q$, the distribution of orientation becomes narrower, as reflected by the increase in ALF. Here again, we can observe the asymmetry of H-bonds: while the position of the maximum of $|P_\parallel|$ is almost unchanged when varying $\Delta q$, two configurations with a similar values of ALF (see the insets for $\Delta q=0.18e$ and $0.42e$), the corresponding values of $|P_\parallel|$ is larger when the molecule receives a H-bond from the octanol at this position than in the reverse case.


\subsection{Clay surfaces \label{subsec:clays}}

\begin{figure}[h!]
    \centering
    \includegraphics[width=0.45\columnwidth]{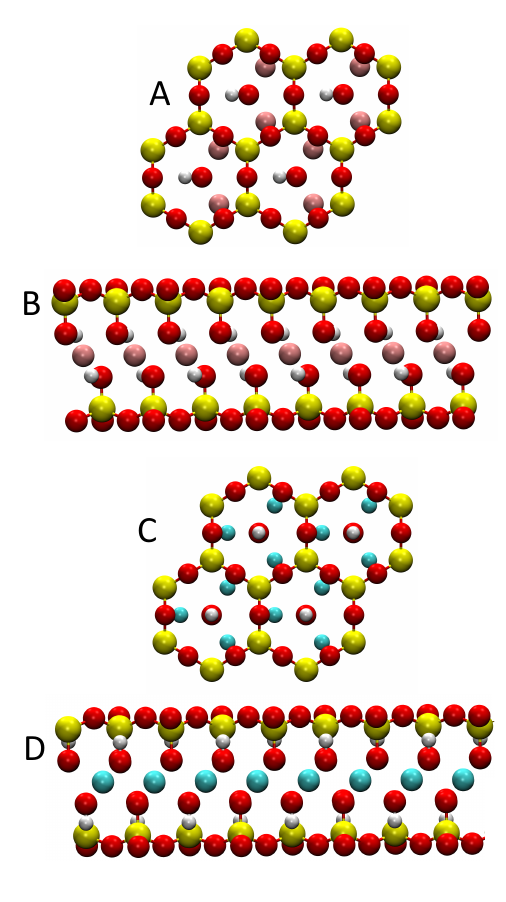}
    \caption{
    Top  and side views of pyrophyllite (A,B) and talc (C,D). Oxygen atoms are in red, hydrogen atoms in white, silicon atoms in yellow, aluminium atoms in pink and magnesium atoms in blue. Fluorotalc has the same structure as talc except that the hydroxyl groups in the hexagonal cavities are replaced by fluorine atoms. }
    \label{fig:argile_view}
\end{figure}

After having considered two molecular solutes, we now turn to the study of interfaces by focusing on three uncharged clay-like minerals namely talc, fluorotalc and pyrophyllite, whose microscopic structures are displayed in Fig.~\ref{fig:argile_view}. Each clay sheet is composed of an octahedral layer of magnesium or aluminium oxide sandwiched between two tetrahedral layers of silicon oxide. The silicon oxide sheet has a honeycomb-like structure. All the octahedral positions are occupied by magnesium in talc and fluorotalc. In pyrophyllite, only two-thirds of the octahedral sites are occupied by aluminium cations. The clays are globally neutral since the positive charges of {Mg$^{2+}$} and {Al$^{3+}$} are neutralized by the silicate groups forming the tetrahedral layer and by hydroxyl groups located at the centers of the hexagonal rings. In talc, these hydroxyls are perpendicular to the surface, they  are parallel to the surface in pyrophyllite while they are replaced by fluorine atoms in fluorotalc. Therefore, the hydroxyl can form hydrogen bonds with water molecules in talc.


In order to study the interfaces between liquid water  and these clay mineral surfaces, we consider a $L_x\times L_y\times L_z$ box with $L_z=30$~\AA\ for all clays and $L_x=10.48$~\AA, $L_y=18.19$~\AA\ for talc and fluorotalc, while $L_x=20.72$~\AA\ and $L_y=35.88$~\AA\ for pyrophyllite. These dimensions correspond to ${42}\times{73}$ and ${82}\times{144}$ unit cells in the $x$ and $y$ directions, respectively. The grid resolution is $\approx0.25$~\AA\ in these directions.  Since our focus is on projections along the $z$ direction, a finer grid resolution of 0.1~\AA\ was employed for $L_z$. Here again, 196 discrete orientations are used to describe the rotation space spanned by the three Euler angles. 

\begin{figure}[ht!]
    \centering
    \includegraphics[width=0.8\columnwidth]{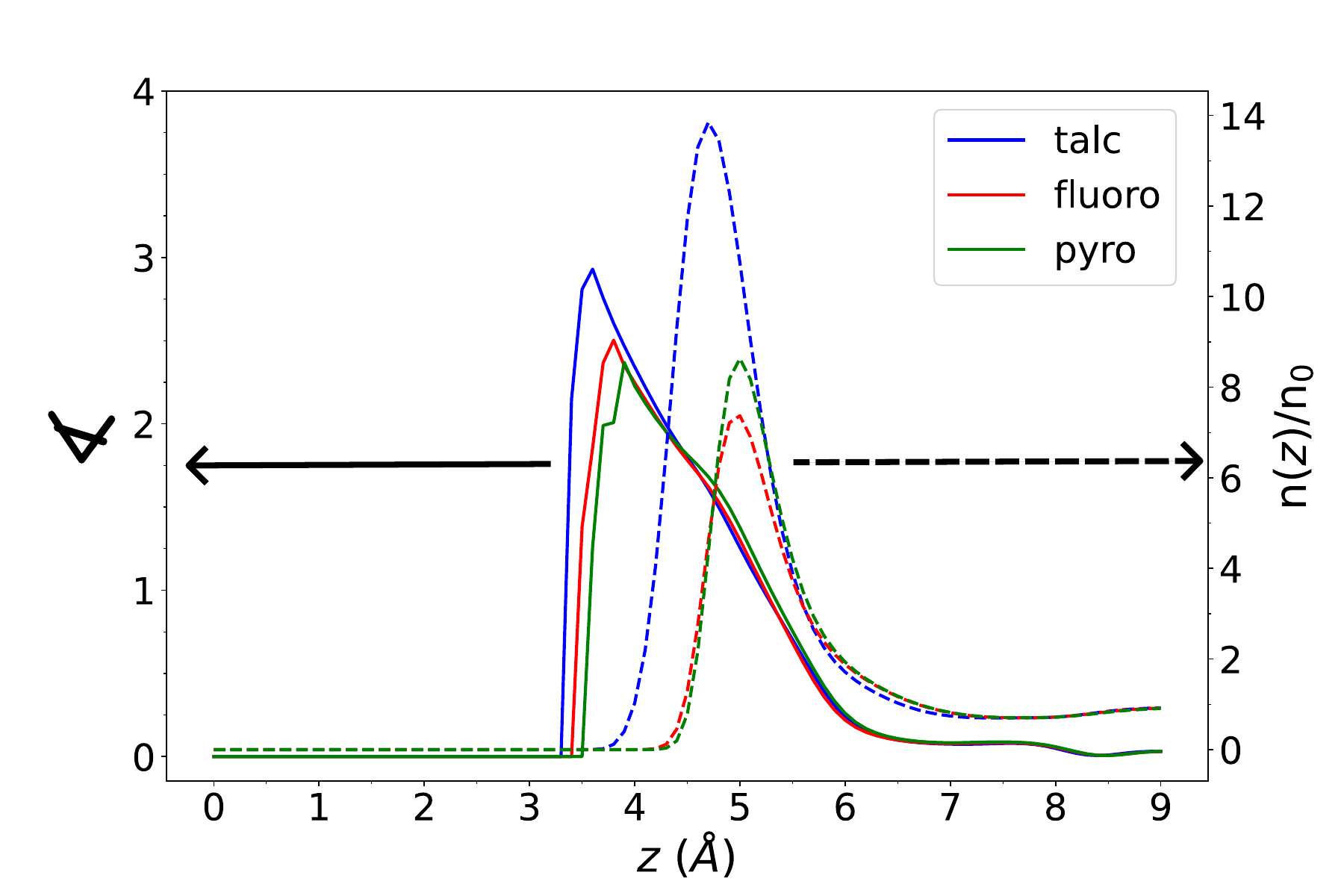}
    \caption{
    Angular localization function (solid lines) and oxygen density (dashed lines) along the line perpendicular to the surface and passing through the center of the hexagonal rings. The curves for pyrophyllite, talc and fluorotalc are shown in green, blue and red, respectively.}
    \label{fig:Alf_dens_argiles}
\end{figure}

The number density profiles along the direction $z$ perpendicular to the surface and passing through the center of the hexagonal ring are reported for the three clay surfaces in Fig.~\ref{fig:Alf_dens_argiles}. The origin of the $z$ axis is located in the central plane of the clay sheet, containing the the Mg$^{2+}$ or Al$^{3+}$ ions. The density profiles of the fluorotalc and pyrophyllites are quite similar and consistent with the potentials of mean force (PMF) at saturation (\textit{i.e.} at a chemical potential corresponding to the liquid-vapor equilibrium of water) reported for the same clays and force fields~\citen{rotenberg_molecular_2011}. The density maximum, located at $z=5.1$~\AA\, is slightly closer to the surface than expected from the PMF which exhibits a minimum at 6.1~\AA.  The density profile for talc differs from the other two, with a more pronounced maximum that is shifted closer to the surface.  This is also in qualitative agreement with the  PMF reported in Ref.~\citenum{rotenberg_molecular_2011}, which exhibits an additional local minimum above the binding site. Note that the density profiles presented in Fig.~\ref{fig:Alf_dens_argiles} exhibit a single peak whereas the PMF for  talc reported in Ref.~\citenum{rotenberg_molecular_2011} exhibits two minima, suggesting the presence of two distinct density peaks.  This is because in the latter case the PMF were computed by averaging over the $x$ and $y$ directions, while the  density profiles reported in Fig.~\ref{fig:Alf_dens_argiles} are calculated along a specific line passing through the cavity. When the density predicted by MDFT is also averaged over the $x$ and $y$ directions, a primary peak preceded by a smaller secondary peak (or shoulder) are present, as previously reported for  pyrophyllite.\citen{jeanmairet_hydration_2014}. The density peaks reported in  Fig.~\ref{fig:Alf_dens_argiles} correspond to the prepeaks (or shoulders ) of the plane averaged densities.

Fig.~\ref{fig:Alf_dens_argiles} also reports the ALF profiles in the same direction. The main feature of all three profiles is a broad peak, with a maximum closer to the surface than the one for density followed by a shoulder at the position of the latter. The shoulder indicates that the molecules in the first density peak are not isotropically oriented. In general, the structure of water at the surface of clays results from a subtle balance between H bonds with the mineral surface, as well as H bonds received from and donated to other water molecules (see \textit{e.g.} Ref.~\citenum{marry_structure_2008}). The maximum of ALF corresponds however to molecules that are closer to the surface, \textit{i.e.} inside the hexagonal cavity, where water molecules preferentially adsorb at low relative humidity -- unlike at saturation due to the balance between water-surface and water-water interactions~\citen{rotenberg_molecular_2011}. In such configuration, water molecules adopt specific orientations, with the oxygen inside the cavity and the dipole pointing away from the surface. The maximum in ALF reflects this strong deviation from the isotropic distribution of orientations. As for the number density, we also observe that talc behaves differently from the other two clay minerals, with a maximum that is larger and for a smaller $z$ value. This is due to the orientation of the hydroxyl group inside the cavity, which can donate a H bond to the water molecule inside the cavity, with a stronger local electric field in the $z$ direction compared to the case of pyrophyllite (where the hydroxyl is tilted with respect to the $z$ direction) and fluorotalc (where the hydroxyl is replaced by fluorine). Note that all the above discussion of the orientational order is made possible by considering ALF, whereas the orientational entropy $S_\Omega$ fails to highlight these features due to the small number of molecules present in the cavities under saturation conditions.

\section{Conclusions}

In this work, we introduced the Angular Localization Function (ALF),  a tool to quantify and visualize the local orientational order of solvent molecules in inhomogeneous fluids. Inspired by the Electron Localization Function (ELF) in electronic density functional theory, ALF measures the local deviation of the angular distribution of molecules from an isotropic reference state. Mathematically, ALF is the Kullback-Leibler divergence between the isotropic distribution of orientations and the considered one, but it can also be naturally derived from the ideal part of the functional.

We demonstrated the utility of ALF by applying it to three distinct systems immersed in water: a water molecule as a solute, an octanol molecule, and three neutral clay minerals (talc, fluorotalc, and pyrophyllite). These case studies revealed that ALF provides complementary insights to traditional metrics such as number density or average dipolar polarization. For instance, ALF effectively highlights regions where solvent molecules exhibit strong orientational preferences, even in low-density areas, which are often challenging to capture with conventional methods. Around a water molecule or octanol, ALF identified regions of high anisotropy near hydrogen-bonding sites, while at clay interfaces, it characterized the fine organization of water molecules within hexagonal cavities. 

A key advantage of ALF lies in its visualization capabilities, making it a valuable tool for analyzing complex solvation structures, such as those encountered near biological molecules or mineral surfaces. Unlike molecular simulation methods, which can suffer from sampling limitations, ALF can be efficiently computed within the framework of Molecular Density Functional Theory (MDFT), significantly reducing computational costs while maintaining high spatial and angular resolution. We believe ALF will be particularly helpful to study systems where orientational order plays a critical role, such as solid-liquid interfaces or solvation environments around macromolecules. 
In summary, ALF is a powerful tool to analyse the solvent structure, adding a new dimension to the interpretation of data obtained from MDFT or molecular simulations.

\section*{Supporting Information}
\gj{Cartesian coordinates , partial charges  and Lennard-Jones parameters  for the octanol molecule are reported in SI.}

\section*{Acknowledgments}

This work was supported  by the France 2030 program, project RADICAL (Grant ANR-23-PEBA-0005). This project received funding from the European Research Council under the European Union’s Horizon 2020 research and innovation program (grant agreement no. 863473). The authors acknowledge the use of Le Chat (Mistral AI) to refine the English language when preparing this manuscript. To readers familiar with the namesake TV show, the authors wish to assure that no cats were harmed by the introduction of ALF in the present work.

\section*{Author declarations}

\subsection*{Conflict of interest}
There is no conflict of interest to declare.

\subsection*{Author contributions}

\textbf{Ma\"iwenn Sou\^etre:} Conceptualization (supporting); Formal analysis (equal); Investigation (lead); Methodology (supporting); Software (supporting); Writing/Original Draft Preparation (supporting); Writing – review \& editing (equal). \textbf{Benjamin Rotenberg:} Conceptualization (equal); Formal analysis (equal); Funding Acquisition (equal); Investigation (supporting); Methodology (equal); Supervision (supporting); Writing – review \& editing (equal). \textbf{Guillaume Jeanmairet:} Conceptualization (equal); Formal analysis (equal); Funding Acquisition (equal); Investigation (supporting); Methodology (equal); Software (lead); Supervision (lead); Writing/Original Draft Preparation (lead); Writing – review \& editing (equal).

\section*{Data availability}

The original data presented in this study are openly available in Zenodo at 
[DOI/URL] inserted after final acceptance.

\section*{References}

\bibliographystyle{unsrtnat}  
\bibliography{Alf}

\end{document}